\documentclass[runningheads]{svmult}

\usepackage{makeidx}   % allows index generation
\usepackage{graphicx}  % standard LaTeX graphics tool
\usepackage{epsfig}
\usepackage{subeqnar}  % subnumbers individual equations
\usepackage{multicol}  % used for the two-column index
\usepackage{physprbb}  % modified textarea for proceedings,
\makeindex             % used for the subject index
\begin{document}
\title*{H-alpha Stacked Images Reveal\protect\newline Large Numbers of PNe in the LMC}
\toctitle{H-alpha stacked images reveal large
numbers\protect\newline of PNe in the LMC}
% allows explicit linebreak for the table of content
%
%
\titlerunning{Discovery of New PNe using H-alpha stacked images}
% allows abbreviation of title, if the full title is too long
% to fit in the running head
%
\author{Warren Reid\inst{1}
\and Quentin Parker\inst{1,2} }
\authorrunning{Warren Reid et al.}
% if there are more than two authors,
% please abbreviate author list for running head
%
%
\institute{Macquarie University, Sydney, Australia \and
Anglo-Australian Observatory, Sydney, Australia}

\maketitle              % typesets the title of the contribution

\begin{abstract}
Our new, deep, high resolution H$\alpha$ and matching R-band UKST
multi-exposure stack of the central 25 sq. degrees of the LMC
promises to provide an unprecedented homogeneous sample of
$>$1,000 new PNe. Our preliminary 2dF spectroscopy on the AAT has
vindicated our selection process and confirmed 136 new PNe and 57
emission-line stars out of a sample of 263 candidate sources
within an initial 2.5 sq. deg. area. To date approximately one
third of the entire LMC has been scanned for candidates ($\sim$7.5
sq. deg.). More than 750 new emission sources have been catalogued
so far along with independent re-identification of all known and
possible PNe found from other surveys.

Once our image analysis is complete, we plan comprehensive
spectroscopic follow-up of the whole sample, not only to confirm
our PN candidates but also to derive nebula temperatures and
densities which, with the aid of photoionization modeling, will
yield stellar parameters which are vital for constructing H-R
diagrams for these objects. A prime objective of the survey is to
produce a Luminosity Function which will be the most accurate and
comprehensive ever derived in terms of numbers, magnitude range
and evolutionary state; offering significant new insights into the
LMC's evolutionary history. The observation and measurement of our
newly discovered AGB halos around 60\% of these PN will also
assist in determining the initial- to final-mass ratios for this
phase of stellar evolution.

\end{abstract}

\section{Background}
Since Henizes (1956) H$\alpha$ survey of the Magellanic Clouds,
subsequent H$\alpha$ surveys have gone progressively deeper by
increasing resolution and improving observational configurations.
Although this resulted in improved sensitivity and increased the
number of emission-line detections, observers continued to use
single objective-prism plates and separated emission-line stars
from nebulae according to whether a continuum could be seen
adjacent to the H$\alpha$ line.
%The deeper surveys moreover, do not cover the ~250 sq. deg. field
%occupied by the LMC.

The dominant method for the discovery of LMC PNe has been the
identification of the $[OIII]$ 5007, 4959 lines on objective-prism
plates.  The high number of candidates rejected through follow-up
spectroscopy (Morgan 1995) shows that secondary plate images may
play an important role in eliminating spurious identifications
prior to spectroscopy. The need for a thorough, deep, H$\alpha$
survey of the LMC to detect PNe became clear. As part of the
AAO/UKST H$\alpha$ survey of the South galactic plane (Parker \&
Phillips 2003), an equivalent mini-survey of the entire LMC and
surrounding regions was also undertaken. This included an
additional 12x 2-hour exposures and 6x 15-minute exposures on the
central LMC field. The intension was to form a multi-exposure
stack of these exposures to gain significant additional depth in
order to search for faint emission-line sources.

\subsection{The H-alpha Filter}
The filter is effectively the world's largest monolithic
interference filter to be used in Astronomy. The central
wavelength of 6590\AA\ and bandpass 70\AA\ work effectively in the
UKST's fast f/2.48 converging beam (305mm dia gives a 5.5deg.
field. Peak transmission 85\%.) (Parker \& Bland-Hawthorn, 1998).

\subsection{The imaging detector: Kodak, Tech-Pan emulsion}
Generally speaking, in good seeing conditions, Tech-Pan UKST R-
band exposures go about 1 magnitude deeper than the equivalent
standard IIIaF R band images with better imaging, improved
resolution and lower noise characteristics. Both the H$\alpha$ and
contemporaneous red exposures were made on Tech-Pan emulsion. The
use of the same emulsion for both H$\alpha$ and SR exposures
ensures an excellent correspondence of their image point spread
functions when film pairs are taken under the same observing
conditions (Parker and Phillipps 2001). The Schmidt Tech-Pan films
(e.g. Parker \& Malin 1999) are ideal tools to find large numbers
of PNe candidates within the LMC due to their inherent high
resolution.

\subsection{Data Reduction}
The SuperCOSMOS machine in Edinburgh scanned and co-added the
individual H$\alpha$/SR exposures on a pixel grid creating 0.67
arcsec pixels after 15-bit digitized plate scanning. The exposures
reach depths of $\sim$21.5 (4.5 x 10$^{-15}$cm$^{2}$ s$^{-1}$ \AA)
for broad-band Red and $\sim$22 for the equivalent H$\alpha$. This
is much deeper than previous LMC H$\alpha$ surveys such as Lindsay
\& Mullan (1963) R$\sim$15 and Bohannan \& Epps (1974)
R$\sim$14.5. Other more recent surveys have touched on the
Magellanic Clouds in the process of surveying both hemispheres.
The most contemporary of these is Southern H-alpha survey (SHASSA)
undertaken at CTIO (Gaustad et.al., 2001). This is a full
hemisphere survey however at 48 arcsec resolution, it doesn't
offer the level of spacial resolution provided by the UKST H-alpha
survey. The UM/CTIO Magellanic Cloud Emission-line Survey
conducted by Chris Smith and the MCELS team (1998) uses a CCD
detector on a Schmidt telescope covering the central 8 sq. deg. of
the LMC. The instrument provides 2.035$^{\prime}$ pixel$^{-1}$,
giving $\sim$3$^{\prime\prime}$ resolution with a field of view of
1.1 sq. deg.. Only the UKST survey can therefore detect and
identify large numbers of faint point source emitters.
 %It is also
%deeper than the previous H$\alpha$ survey conducted on the UKST in
%1979  by Davies, Eliot and Meaburn (DEM) which used a 105$\AA$
%band width filter and Kodak 098-04 emulsion, which they combined
%with 103 emoltion giving band width of 80$\AA$).
Bland-Hawthorn, Shopbell and Malin (1993) have shown that digital
stacking of UKST plates and films can achieve canonical poissonian
depth gains. We estimate our stacked images reach $\sim$1.35
magnitudes deeper for H$\alpha$ and 1 magnitudes fainter in Red
than an individual exposure.

Another significant advantage of the stacking process is that
small emulsion imperfections and adhering dust particles,
scratches etc which can be problematic and lead to spurious
detections are naturally eliminated as part of the combination
process. Likewise the influence of variable stars is considerably
alleviated by the time averaging process by stacking multiple
exposures taken over a three year period.

%The SuperCOSMOS processing also produces the Image Analysis Mode
%or IAM which simply presents 32 parameters for each object
%detected down to about the second faintest magnitude. (Ra, Dec,
%brightest intensity above sky, left extent, right extent, total
%area, magnitude, unweighted semi-major axis, (minor axis),
%classification flag, celestial position angle, de-blending flag,
%area above areal profile for 8 levels)

\section{Detection Technique}

Candidate emission sources are found using an adaptation of a
technique whereby fits images from the deep SR stack are first
coloured red and overlaid upon the matching stacked H$\alpha$
images coloured blue. The coloured images are then merged and
matched for point spread function (psf) which is achieved both
visually and through numerical intensity values, constantly
displayed in the RGB program of KARMA.
%Background and star matching are achieved by adjusting the
%intensity level for each displayed image until the levels are
%equal for each of the overlapping H$\alpha$ and R images when
%centered on either ordinary main-sequence stars or 'empty'
%background.
All continuum sources such as LMC Stars and background galaxies
become a uniform colour but emission nebulosities and candidate
compact emission sources develop a blue hue, making discovery and
identification straightforward. Careful selection of software
parameters allows the intensity of the matching H$\alpha$ and SR
images to be perfectly balanced allowing only peculiarities of one
or other band-pass to be observed and measured. $\textbf{All}$
previously known LMC PNe have been re-identified either by their
extended halos surrounding the central PN shells or as areas of
compact dense emission. In each case the H$\alpha$ emission is
seen as a bright blue aura.

Our preliminary application of this technique, applied to several
1-square degree sub-regions of the main LMC field, has produced
extremely encouraging results. Areas were chosen which contained
known emission sources and PNe, including some of the previous
faintest detected sources such as PNeJ07 at m$_{B}$21.7, which was
easily identified. Scrutiny of the combined, KARMA processed,
1-degree H$\alpha$/SR images revealed dozens of new PNe candidates
in each sub-field.
%As an example, in a 1-deg field centered on (J2000) 05:45:50
%-70:48:51, we identified all 5 known PNe in the region and
%revealed a 114 further emission sources. Of these, 3 have been
%previously catalogued as emission objects of unknown nature, five
%have been catalogued as small HII regions and one as an SNR. The
%remaining 92 emission sources have no catalogue references. Seven
%small bubble-type rings of H$\alpha$ emission ( approx. 3 arcsec
%dia.) have also been detected when no central star is visible in
%the matching SR image. Magnitudes of these sources in m$_{B}$
%reach down to 22.73. At a distance of 51 Kpc, these objects span a
%diameter of 0.73pc; not untypical for evolved PNe.
Several point sources exhibit very centrally concentrated
emission, while others display rings and faint outer bubbles,
halos or extensions. The enhanced angular widths of many of the
emission sources revealed by our technique (up to  4 arcseconds in
radius about the central star) strongly favour PNe in these cases.

\section{Identification of new candidate LMC emission sources}

Only spectroscopic data will unambiguously determine the nature of
the new candidate sources whilst their magnitudes (15$\leq$ R
$\leq$ 21) and number density (100 sq/deg) make the 2dF
spectrograph on the AAT the obvious choice for effective
follow-up. All previously known PNe in the preliminary studied
regions were successfully re-identified whilst the characteristics
of the newly identified sources continue their trend to fainter
magnitudes. We are including every strong source that exhibits a
halo whose diameter is $>$20\% the diameter of the central source
regardless of the fact that many of them will be emission-line
stars. The fact that Galactic Bulge Symbiotic stars and compact
PNe look the same means we will not discard any star that appears
to exhibit H$\alpha$ emission. Therefore, we can be confident that
we are extracting every unidentified PNe to the limit of the
stacked data. Figure 1 shows the results of an initial AAT 2dF
service time run where we were able to place almost 2/3 of the
sources in a 1.25 square degree area onto optical fibres for
spectral confirmation. It can be seen that our data is extending
the LMC PNe range both in terms of numbers of detections and depth
of magnitude.

The apparent visual density of the H$\alpha$ emission has proven
to be a fair indication of the source type. All known PN are
strong radiative emitters with thick emission halos that are more
than double the visual diameter of the central source. By
comparison with HST imaging of LMC PNe, we find film saturation
increasing the diameter of the central PN source including shell
structure by $\sim$4 times for an average magnitude 16 central PN
at logarithmic intensity levels. This figure decreases to $\sim$2
times for the widest, large structure PNe such as SMP 93. Faint,
wide- scale structures such as the AGB halos surrounding PNe will
not suffer from the same PSF characteristics so we can be more
confident about the large diameters we are observing.

As candidates are discovered, we record position, check data bases
including (but not restricted to) SIMBAD for previous detections,
allot an ID number and give each a probability rating. A note is
made relating to any peculiarities associated with the source such
as nearby bright or overlapping stars, bi-polar emission, shape,
intensity, optical diameter, density, and proximity of the
emission to any nearby $HII$ regions.  Magnitudes are found with
the STARLINK PHOTOM package which has been calibrated to several
catalogues and the SuperCOSMOS on-line UKST R-band catalogue.

%We have divided the LMC into 16 images each 1° 25' square. The
%superCOSMOS machine has provided Image Analysis Mode (IAM) data
%for each source.  Using the CURSA package provided by starlink,
%the IAM file covering the entire LMC field was subdivided into 16
%segments each exactly matching the co-ordinates of each of the 16
%images covering the same field. Each source was checked in GAIA
%(%starlink software) for blending. Each of the 32 parameters
%associated with that source have then been recorded against that
%source in the data base. COSMAG (parameter 9) is calculated from
%the sky-subtracted intensity values in the connected pixels
%comprising an image and is given in units of millimag. COSMAG is
%an isophotal magnitude. Details may be found in Hambly et al
%(2001).

%Figure 2 is SMP93 with it's AGB halo. The halo is 13.9 arcsec
%along it's major axis which is 3.37 parsecs across. The central PN
%however is also extremely large at approximately 4.2 x 3 arcsec.
%The figure shows where it fits to scale within the UKST image,
%revealing the extent of the halo.

\section{Future Plans}

With such a large number of PNe at relatively the same 50Kpc
distance, the primary objective will be the creation of a
comprehensive luminosity function. This will be achieved for the
H$\alpha$ line to allow us to better understand the evolutionary
types in the LMC. It will also provide a probe into the structure
and origin of the LMC as well as providing a distance probe when
applied to other Hubble-type galaxies. Spectroscopic analysis will
add further to our knowledge of AGB phase evolution as well as
chemical enrichment in the LMC, especially when that information
is combined with morphology and evolutionary types. HR Diagrams
are useful tools for examining the evolution of stars and
understanding the different phases that evolve from differing
stellar and nebula parameters. Extended late AGB phase Halos,
which we have discovered for the first time in LMC PNe, will be
examined, particularly for their diameters and shapes with
relation to the central PNe. The initial- to final-mass relation
is another important area of stellar physics that can now be
re-examined by virtue of measuring halo parameters; since all of
our sources are at the same general distance. Our data-base will
provide a comprehensive catalogue which will be published and
on-line for general access.

\section{Conclusion}

So far 6 out of 16 image cells have been analyzed. The detection
rate increases 3-fold on the main bar. Approximately half the 263
candidates followed-up on 2dF have been confirmed as PN, 57 are
emission-line stars, 33 require longer exposure times, while 33
are non-PNe detections which include late-type stars. At these
rates, we expect to confirm more than 1,000 new PNe in December
when we return to 2dF on the AAT. Also in December we will be
using VLT FLAMES to observe very faint candidates and gain faint
diagnostic lines from a large sample covering a wide evolutionary
range and various positions on and off the main bar.

%\section{Figures}
\begin{figure}[b]
\begin{center}
\includegraphics[width=.9\textwidth]{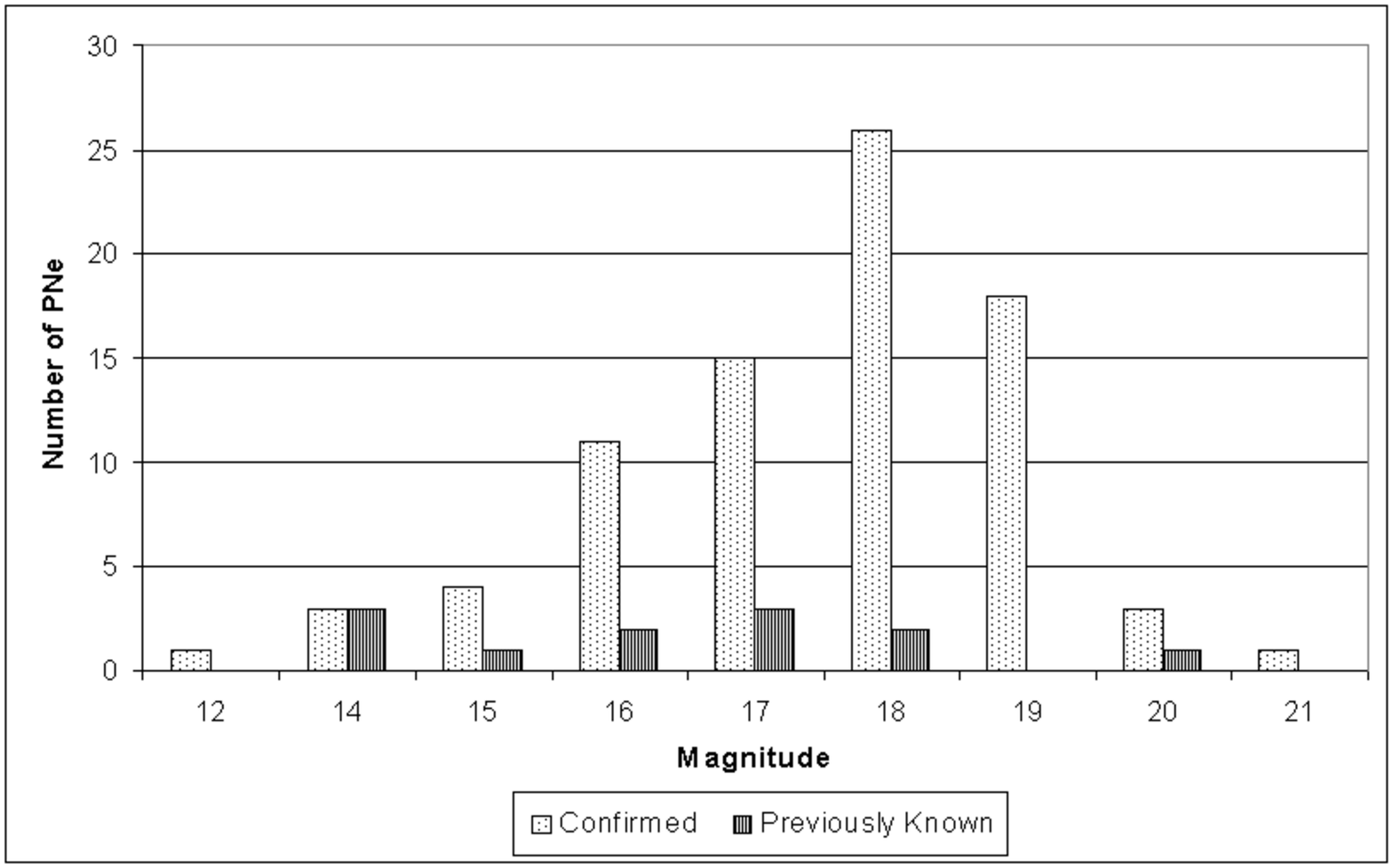}
\end{center}
\caption[]{Results of spectroscopic confirmation in a 1.25 sq.
deg. area; number vs. magnitude} \label{eps1}
\end{figure}

%\section{Figures}
\begin{figure}[b]
\begin{center}
\includegraphics[width=.9\textwidth]{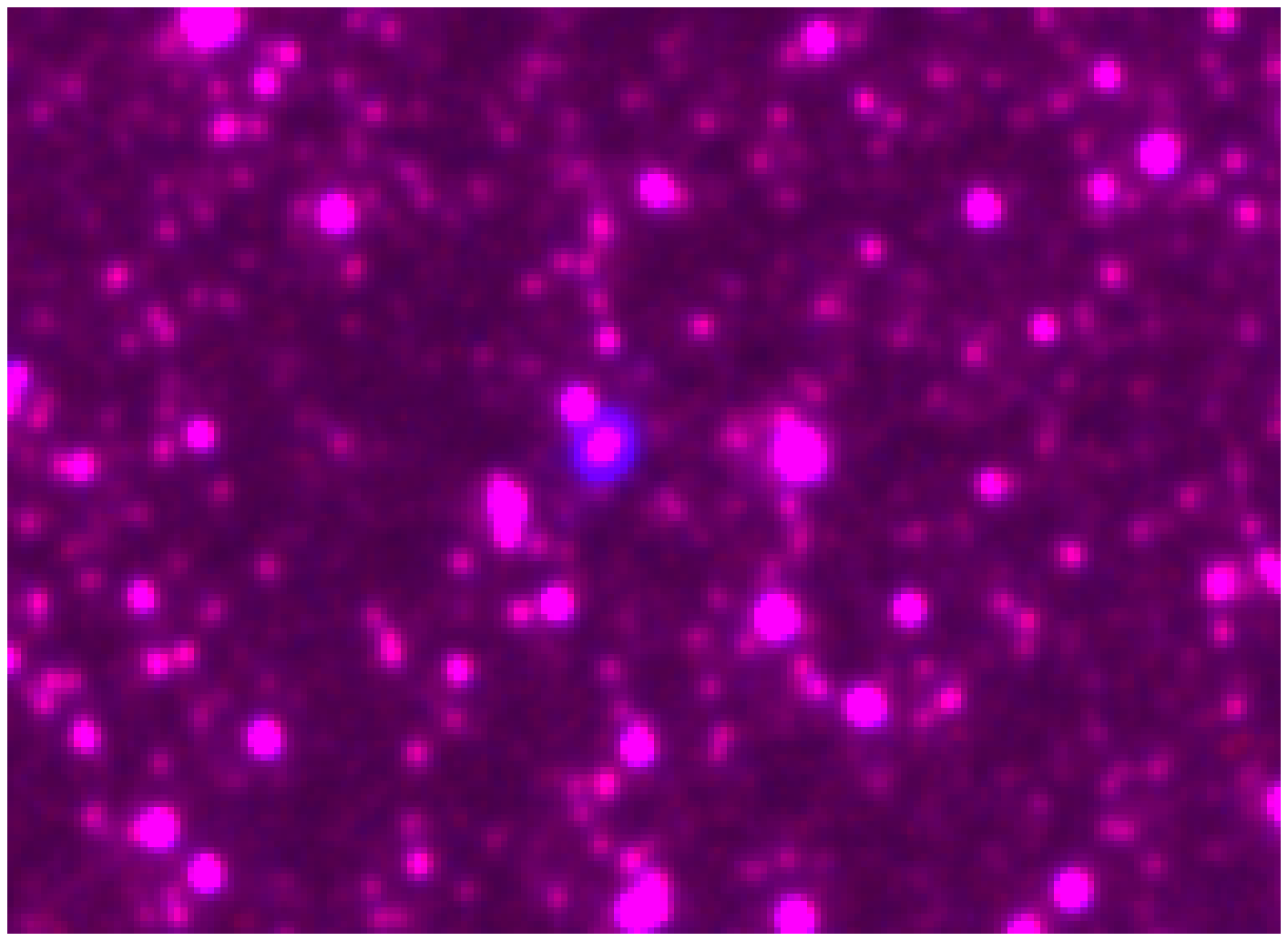}
%\end{flushleft}
\caption[width=.5\textwidth]{Combined colour image of a newly
discovered PN}
\end{center}
\label{eps2}
%\end{figure}

%\section{Figures}
%\begin{figure}[b]
\begin{center}
\includegraphics[width=.9\textwidth]{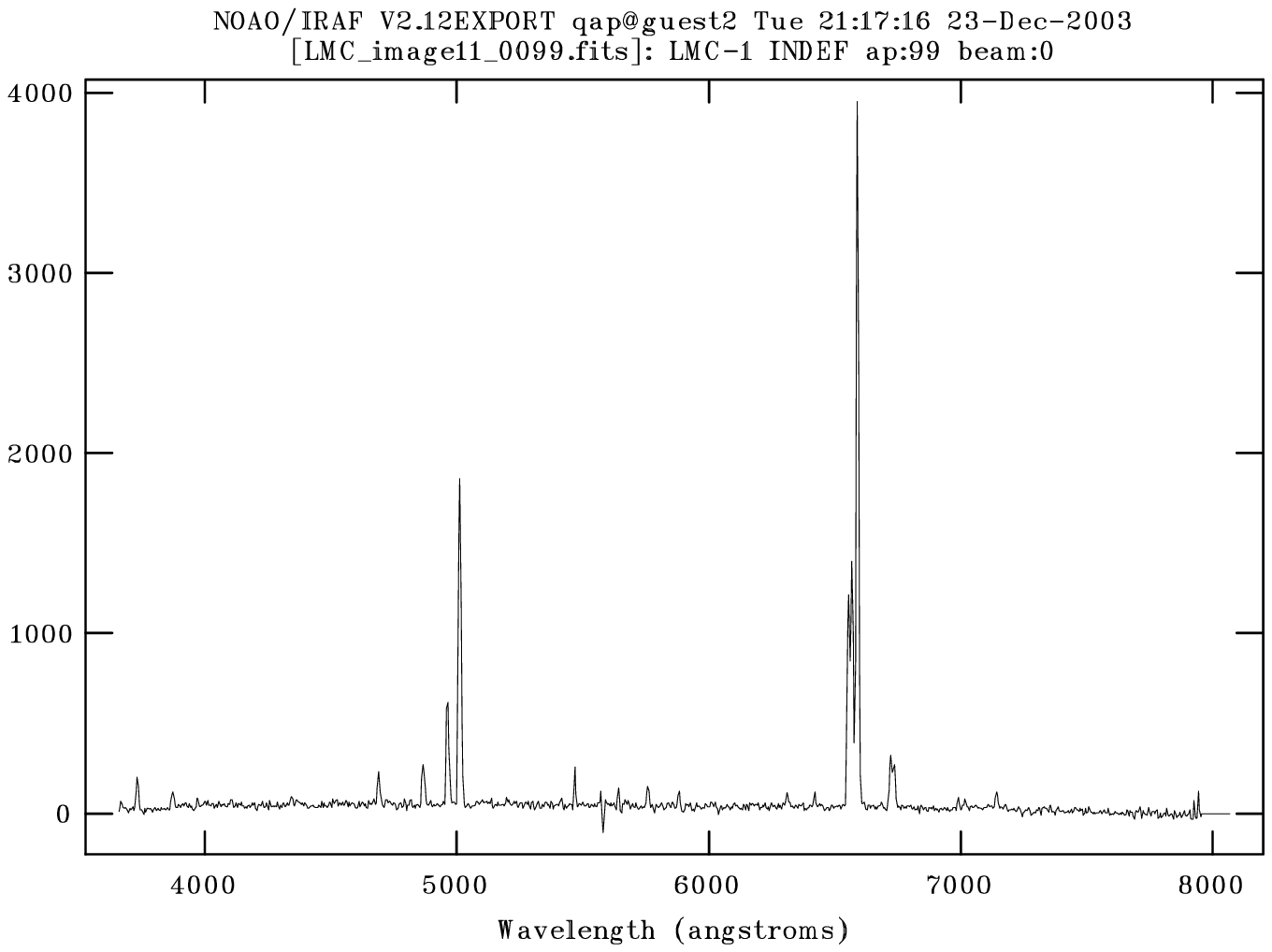}
%\end{flushright}
\caption[width=.5textwidth]{Spectroscopic confirmation}
\label{eps3}
\end{center}
\end{figure}

%In order to mark the desired amount of space for a (centered!)
%figure which has to be pasted into the manuscript manually please
%provide a vertical line on the lefthand side of the figure. This
%is reached by using the commands \verb|\mpicplace{width in
%cm}{height in cm}|.

%INDEX%%%%%%%%%%%%%%%%%%%%%%%%%%%%%%%%%%%%%%%%%%%%%%%%%%%%%%%%%%%%%%%
% Please check with the editor of your book whether he plans to
% include a "mutual" subject index - if so, please code your entries
% in the standard syntax. For your own purposes you may print your
% "personal" index by using the following commands:
%
%\clearpage
%\addcontentsline{toc}{section}{Index}
%\flushbottom
%\printindex
%%%%%%%%%%%%%%%%%%%%%%%%%%%%%%%%%%%%%%%%%%%%%%%%%%%%%%%%%%%%%%%%%%%%%

\end{document}